\documentclass[aps,pre,letterpaper,twocolumn,floatfix,showpacs]{revtex4}
\usepackage{graphicx,amsmath}

\newcommand{\Pe}{\mathrm{P\hspace{-1pt}e}}
\newcommand{\Peo}{\mathrm{P\hspace{-1pt}e}\hspace{-0.5pt}_o}
\newcommand{\Gc}{G_{\hspace{-1.5pt}c}}
\newcommand{\Ginf}{G_{\hspace{-1.5pt}\infty}}
\newcommand{\Hth}{H_{\hspace{-0.5pt}\theta}}

\renewcommand{\Re}{\mathrm{Re}\,}

\newcommand{\del}{\boldsymbol{\nabla}}
\newcommand{\nhat}{\mbox{\boldmath{$\hat{n}$}}}
\newcommand{\xhat}{\mbox{\boldmath{$\hat{x}$}}}

\newcommand{\mbs}[1]{\boldsymbol{#1}}
\newcommand{\mbf}[1]{\mathbf{#1}}

\newcommand{\ld}{\lambda}
\newcommand{\LD}{\Lambda}
\renewcommand{\th}{\theta}
\newcommand{\ens}[2][]{\langle{#2}\rangle_{#1}}
\newcommand{\intth}[1][\theta]{\int_0^{2\pi}\hspace{-4pt}d{#1}\;}

\newcommand{\bump}{\phi}

\begin{document}

\title{The Average Shape of Transport-Limited Aggregates}

\author{Benny Davidovitch$^1$, Jaehyuk Choi$^2$, and Martin Z. Bazant$^2$}
\affiliation{
  $^1$ Division of Engineering and Applied Sciences,
  Harvard University,
  Cambridge, MA 02138 \\
  $^2$ Department of Mathematics,
  Massachusetts Institute of Technology,
  Cambridge, MA 02139}

\date{\today}

\begin{abstract}
We study the relation between stochastic and continuous
transport-limited growth models, which generalize conformal-mapping
formulations of diffusion-limited aggregation (DLA) and viscous
fingering, respectively.  We derive a nonlinear integro-differential
equation for the asymptotic shape (average conformal map) of
stochastic aggregates, whose mean-field approximation is the
corresponding continuous equation, where the interface moves at its
local expected velocity.  Our equation accurately describes
advection-diffusion-limited aggregation (ADLA), and, due to nonlinear
averaging over fluctuations, the average ADLA cluster is similar, but
not identical, to an exact solution of the mean-field
dynamics. Similar results should apply to all models in our class,
thus explaining the known discrepancies between average DLA clusters
and viscous fingers in a channel geometry.
\end{abstract}


\maketitle

Developing effective mean field approximations to nonlinear
stochastic equations constitutes a major challenge in various
active fields of statistical physics, e.g. hydrodynamic turbulence
\cite{Frisch} and self organized criticality \cite{Bak}.
Straightforward derivation of such approximate theories typically
involves uncontrolled assumptions required to "close" an infinite
hierarchy of equations for moments of the probability distribution
of the stochastic field. An alternative approach consists of
deriving asymptotic solutions to a deterministic version of the
original stochastic dynamics, assuming that such solutions capture
the behavior of ensemble average of the original stochastic field
\cite{Bak}. Such approach, however, might turn out to be
unreliable as well, since it underestimates the possible effects
of noise on the asymptotic evolution of a stochastic field.

A nontrivial example in which such approach has been advanced over
the last two decades is the fractal morphology of patterns
observed in computer simulations of the celebrated diffusion
limited aggregation (DLA) model \cite{witten81}. Since the
relation between the mathematical formulations of the stochastic
DLA process and the continuous viscous fingering dynamics was
established \cite{Paterson84}, the striking similarity between
patterns observed in both processes has triggered various attempts
to interpret viscous fingering dynamics as a mean field for DLA
\cite{stanley91,arneodo89,barra02,swinney03}.

In this Letter, we study the connection between a broad class of
stochastic transport-limited aggregation processes and their
continuous counterparts~\cite{bazant03}. In our models, growth is
fuelled by nonlinear, non-Laplacian transport processes, such as
advection-diffusion and electrochemical conduction, which satisfy
conformally invariant equations~\cite{bazant04}. Stochastic and
continuous dynamics are defined by generalizing conformal-mapping
formulations of DLA~\cite{hastings98} and viscous
fingering~\cite{polubarinova45,shraiman84}, respectively. We show
that the continuous dynamics is a self-consistent mean-field
approximation of the stochastic dynamics, which, nevertheless,
does not accurately predict the average shape of a random ensemble
of aggregates.

\begin{figure}
  \parbox[t][][b]{0.49\linewidth}{\flushleft
    (a)\\
    \includegraphics[width=\linewidth]{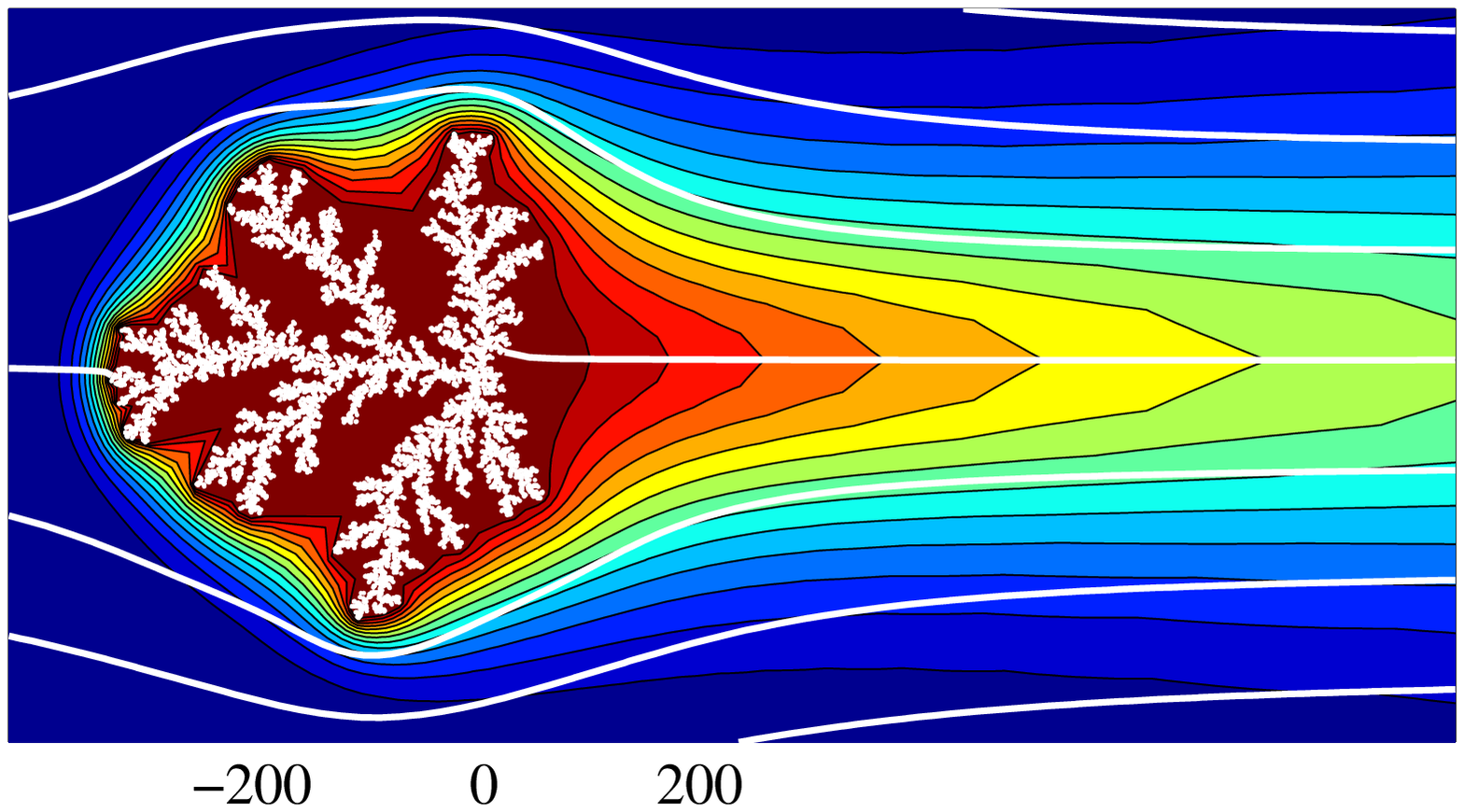}
  }
  \parbox[t][][b]{0.49\linewidth}{\flushleft
    (b)\\
    \includegraphics[width=\linewidth]{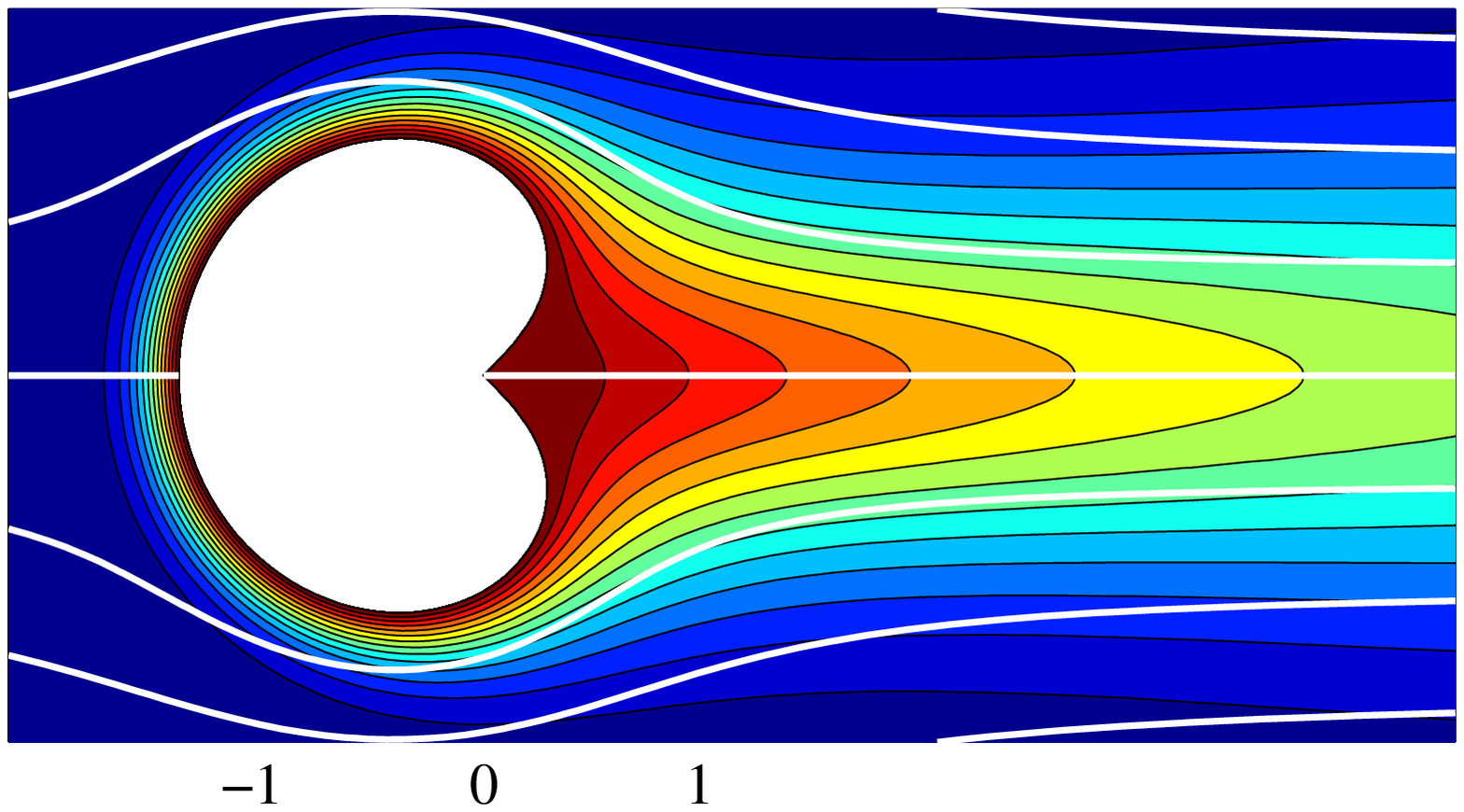}
  }
  \caption{(a) A typical ADLA cluster of $n=10^5$ ``particles''
  (iterated conformal maps). Color contours indicate the particle
  concentration $c$, and solid yellow curves are fluid streamlines.
  (b) The exact, asymptotic shape of the analogous continuous
  dynamics, $\Gc(w)$, which describes solidification in a fluid
  flow. For both (a) and (b), we set $\Pe\approx 20$ so that the
  far fields are similar.}
\end{figure}

We consider a set of two-dimensional scalar fields, $\mbs{\varphi} = \{
\varphi_1, \varphi_2, \ldots, \varphi_M \}$, whose gradients produce
quasi-static, conserved ``flux densities'',
\begin{equation} \label{eq:Fi}
  \mbf{F}_i  = \sum_{j=1}^{M} C_{ij}(\mbs{\varphi}) \del \varphi_j  \;, \qquad
  \del\cdot\mbf{F}_i = 0
\end{equation}
in $\Omega_z(t)$, the exterior of a singly connected domain that
represents a growing aggregate at time $t$.  (The coefficients,
$\{ C_{ij} \}$ may be nonlinear functions of the fields.)  The
crucial property of the nonlinear system (\ref{eq:Fi}) is its
conformal invariance~\cite{bazant04}: If $\mbs{\varphi}
(w,\overline{w})$, not necessarily harmonic, is a solution in a
domain $\Omega_w$ and $w=f(z)$ is a conformal map from $\Omega_z$
to $\Omega_w$, then $\mbs{\varphi} (f(z),\overline{f(z)})$ is a
solution in $\Omega_z$. Using this fact, the evolving domain,
$\Omega_z(t)$, can be described by the conformal map, $z=g(w,t)$,
from the exterior of (say) the unit disk, $\Omega_w$.

Growth is driven by a combination of flux densities, $\mbf{Q} =
\sum_{i=1}^N B_i(\mbs{\varphi}) \mbf{F}_i $, on the boundary with a
local growth rate, $\sigma =
\nhat\cdot\mbf{Q}$, where $\mbs{\hat n}$ is the unit normal
vector at $z\in \partial \Omega_z(t)$.  For continuous, deterministic
growth, each boundary point $z$ moves with a velocity, $\mbs{v}(z) =
\alpha\,\sigma(z)\,\nhat(z)$, where $\alpha$ is a constant. For
discrete, stochastic growth, the initial seed, $\Omega_z(t_0=0)$, is
iteratively advanced by elementary ``bump'' maps representing
particles of area $\ld_0$ at times $t_1, \cdots , t_n$. The waiting
time $t_n - t_{n-1}$ is an exponential random variable with a mean set
by the total integrated flux.  The probability density to add the
$n$th particle in a boundary element $(z,z+dz)
\in \partial \Omega_z(t_{n-1})$ is proportional to $\sigma(z)|dz|$.

The classical models are recovered in the simplest case of one field
($M=1$). DLA corresponds to stochastic growth by diffusion,
$\mbf{F}=\mbf{Q}=-D\del c$, from a distant source ($c\sim\log|z|$ as
$|z|\to\infty$) to an absorbing cluster ($c=0$ for $z\in\Omega_z(t)$),
where $c$ is the particle concentration and $D$ the diffusivity. Viscous
fingering corresponds to continuous growth by the same process, where
$c$ becomes the fluid pressure and $D$ the permeability.

The simplest, nontrivial models with multiple fields ($M=2$)
involve diffusion in a fluid flow. The stochastic case
is advection-diffusion-limited aggregation (ADLA)~\cite{bazant03},
illustrated in Fig. 1a. Particles are deposited around a circular seed
of radius, $L_o$, from potential flow, $\mbs{v} =\nabla \varphi$, of
uniform velocity $U$ far from the aggregate. The dimensionless
transport problem is
\begin{gather} \label{eq:ad}
  \Peo \del \varphi \cdot \del c = \del^2 c, \quad \del^2 \varphi = 0,
  \quad z \in \Omega_z(t) \\
  \label{eq:circbc}
  c=0, \quad \nhat\cdot\del\varphi=0,\quad \sigma = \nhat\cdot\del c,
  \quad z \in \partial\Omega_z(t) \\
  \label{eq:infbc}
  c \to 1, \quad \del\varphi \to \xhat, \quad |z| \to \infty,
\end{gather}
where $c$ is the concentration of particles. Here $x$, $\varphi$, $c$,
and $\sigma$ are in units of $L_o$, $UL_o$, $C$, and $D C/L_o$,
respectively, and $\Peo = UL_o/D$ is the initial P\'eclet
number. Numerical solutions and asymptotic approximations are
discussed in Ref.~\cite{choi04}.

The transport problem is conformally invariant, except for the
boundary condition, Eq.~(\ref{eq:infbc}), which alters the flow speed upon
conformal mapping. Instead, we choose to fix the mapped background
flow and replace $\Peo$ with the renormalized P\'eclet number, $\Pe(t)
= A_1(t)\Peo$, when Eq.~(\ref{eq:ad}) is transformed from
$\Omega_z(t)$ to $\Omega_w$.  The ``conformal radius'', $A_1(t)$, is
the coefficient of $w$ in the Laurent expansion of $g(w,t)$ and scales
with the radius of the growing object~\cite{hastings98,davidovitch99}.
Since $A_1(t)\rightarrow\infty$ for any initial condition, the
flux approaches a self-similar form,
\begin{equation} \label{eq:sigma_infty}
  \sigma(\th; \Pe) \sim 2\sqrt{\Pe/\pi}\; \sin(\th/2)\quad \text{as}
  \quad \Pe\to\infty.
\end{equation}
More generally, there is a universal crossover from DLA ($\sigma =$
constant) to this stable fixed point, where $\Pe(t)=A_1(t)\Peo$ is the
appropriate scaling variable~\cite{choi05}.

The continuous analog of ADLA is a simple model for solidification
from a flowing melt~\cite{kornev94}. More generally, continuous
dynamics in our class is described by a nonlinear equation,
\begin{equation} \label{eq:gsb}
  \Re (\overline{ w \, g'} \; g_t ) = \alpha\; \sigma(w;\Pe(t)) \quad
  \text{for} \quad |w|=1.
\end{equation}
which generalizes the Polubarinova-Galin equation for Laplacian growth
\cite{polubarinova45,shraiman84} ($\sigma=1$). In the case of
advection-diffusion~\cite{kornev94}, only low-$\Pe$ approximations are
known, but we have found an exact high-$\Pe$ solution
of the form, $g(w,t)=A_1(t)\, \Gc(w)$, where
\begin{equation} \label{eq:sol_cont}
  A_1(t) = t^{2/3},\quad \Gc(w) =  w\,\sqrt{1-1/w}.
\end{equation}
This similarity solution to Eq.~(\ref{eq:gsb}) with $\alpha\,
\sigma(\th,t) = \sqrt{A_1(t)}
\sin(\th/2)$ describes the long-time limit, according to
Eq.~(\ref{eq:sigma_infty}). (We do not know the uniqueness or
stability of this solution or whether it can be approached without
singularities from general initial conditions.)  Just as the
Saffman-Taylor finger solution (for $\sigma=1$) has been compared to
DLA in a channel geometry~\cite{somfai_ball02}, this analytical result
begs comparison with ADLA.

As in Ref.~\cite{bazant03}, we grow ADLA clusters by a modified
Hastings-Levitov algorithm~\cite{hastings98}. The random attachment of
the $n$th particle to the cluster is described by perturbing the
boundary  $\partial \Omega_z (t_{n-1}) \to
\partial \Omega_z(t_n)$ by a ``bump'' of characteristic area
$\ld_0$. This leads to the recursive dynamics
\begin{equation} \label{eq:HL}
  g_n(w) = g_{n-1}\circ\bump_{\ld_n,\th_n}(w), \quad g_n(w) = g(w,t_n)
\end{equation}
where $\bump_{\ld,\th}(w)$ is a specific map, conformal in
$\Omega_w$, that slightly distorts the unit circle by a bump of
area $\ld$ around the angle $\th$. The parameter,
$ \ld_n=\ld_0  |g_{n-1}'(e^{i \th_n})|^{-2}$
is the area of the pre-image of such bump under the inverse map
$g^{-1}$. The angle $\th_n$ is chosen with a probability density
$p(\th;\Pe(t_n)) \propto \sigma(e^{i \th};\Pe(t_n))$, so the
expected growth rate is the same as in the continuous dynamics.

For an ensemble of $n$-particle aggregates, a natural definition of
average cluster shape is the conformal map, $\ens{G_n(w)}$,
defined by averaging at a point $w
\in \Omega_w$ all the maps, $G_n(w)=g_n(w)/A_1(t_ n)$, rescaled to have a unit
conformal radius. We then ask: \emph{What is the limiting average
cluster shape, $\ens{\Ginf(w)} = \lim_{n\to\infty}
\ens{G_n(w)}$, and how does it compare to the similarity solution,
$\Gc(w)$, of the continuous growth equation~(\ref{eq:gsb})?} The same
questions apply to any of our transport-limited growth
models, but here we focus on ADLA as a representative case.

To provide numerical evidence, we grow 2000 ADLA clusters of size
$n=10^5$ using the semi-circular bump function in
Ref.~\cite{davidovitch99} (with $a=1/2$).  To reduce fluctuations,
we aggregate small particles, $\ld_0=1/16$, on a large initial
seed ($g_0(w) = w, |w|=1$).  To reach at the asymptotic limit
faster and also match the assumption of $\Gc(w)$, we fix the
angular probability measure, $p_\infty(\th) = \sin(\th/2)/4$
for $\Pe=\infty$, throughout the growth.  In Fig.~2a, we plot the
average contour of the ensemble, $\ens{G_n(e^{i\th})}$ at $n=10^5$
along with that of the continuous solution, $\Gc(e^{i\th})$.  To
give a sense of fluctuations, we also plot a ``cloud'' of points,
$G_n(e^{i\th})$ over uniformly sampled values of $\th$. Fig.~2b
is the zoom-in of the boxed region in Fig.~2a, where we also show
$\ens{G_n(e^{i\th})}$ at $n=10^3$ and $n=10^4$. Although the
convergence of $\ens{G_n(w)}$ is easily extrapolated, the $n=10^5$
line has not reached at the asymptotic limit yet.  The branch
point at $w=1$ seems to be related to the slow convergence.
Ignoring the unconverged area, $\ens{G_n(w)}$ and $\Gc(w)$ are
quite similar, and yet clearly not same. The average,
$\ens{G_n(w)}$, better captures the ensemble morphology reflected
by the cloud pattern than $\Gc(w)$ and the opening angles at the
branch point of the two curves are also different~\cite{choi05}.

\begin{figure}
  \raisebox{-\height}{(a)}
  \raisebox{-\height}{
    \includegraphics[height=0.55\linewidth]{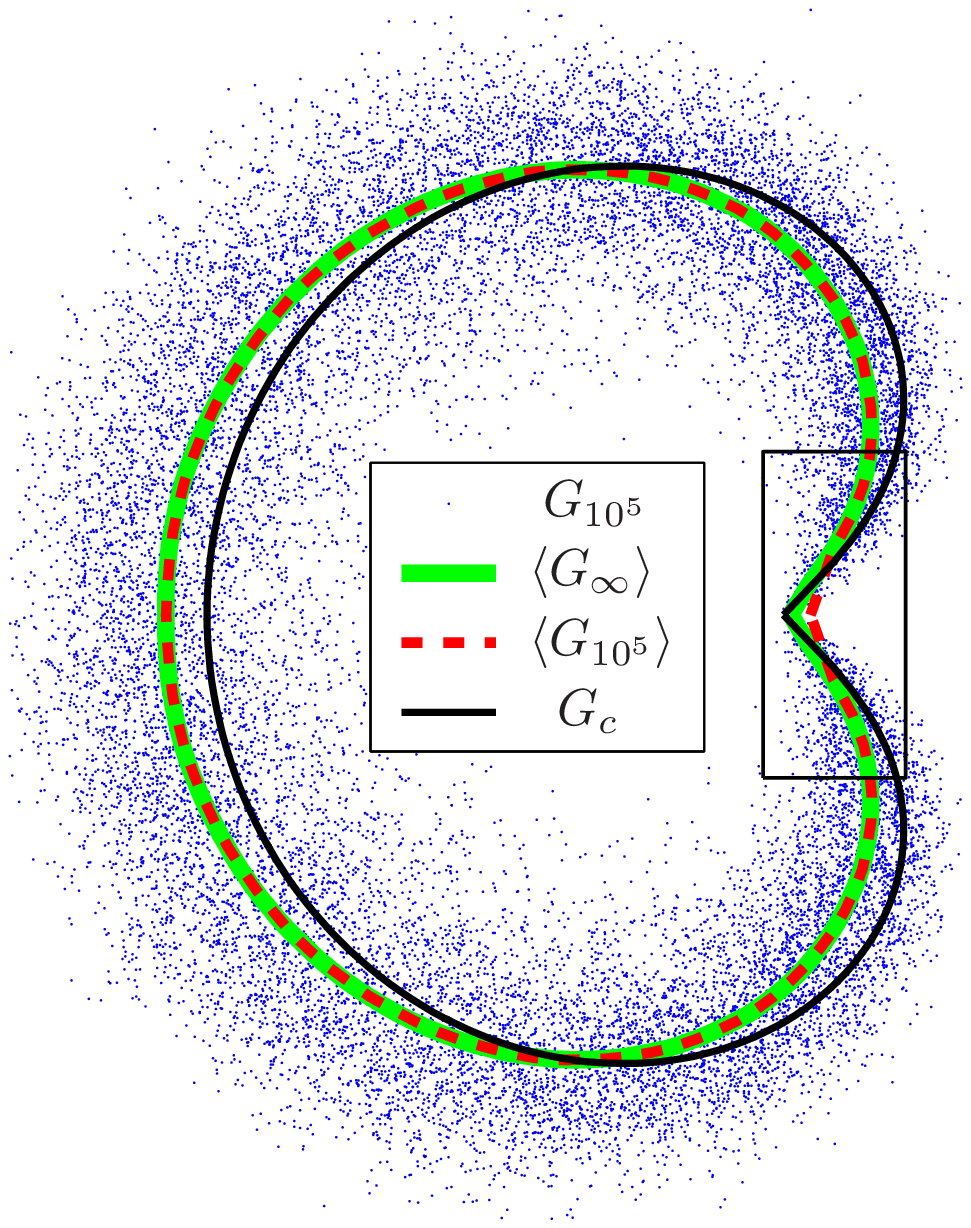}
  }
  \raisebox{-\height}{(b)}
  \raisebox{-\height}{
    \includegraphics[height=0.55\linewidth]{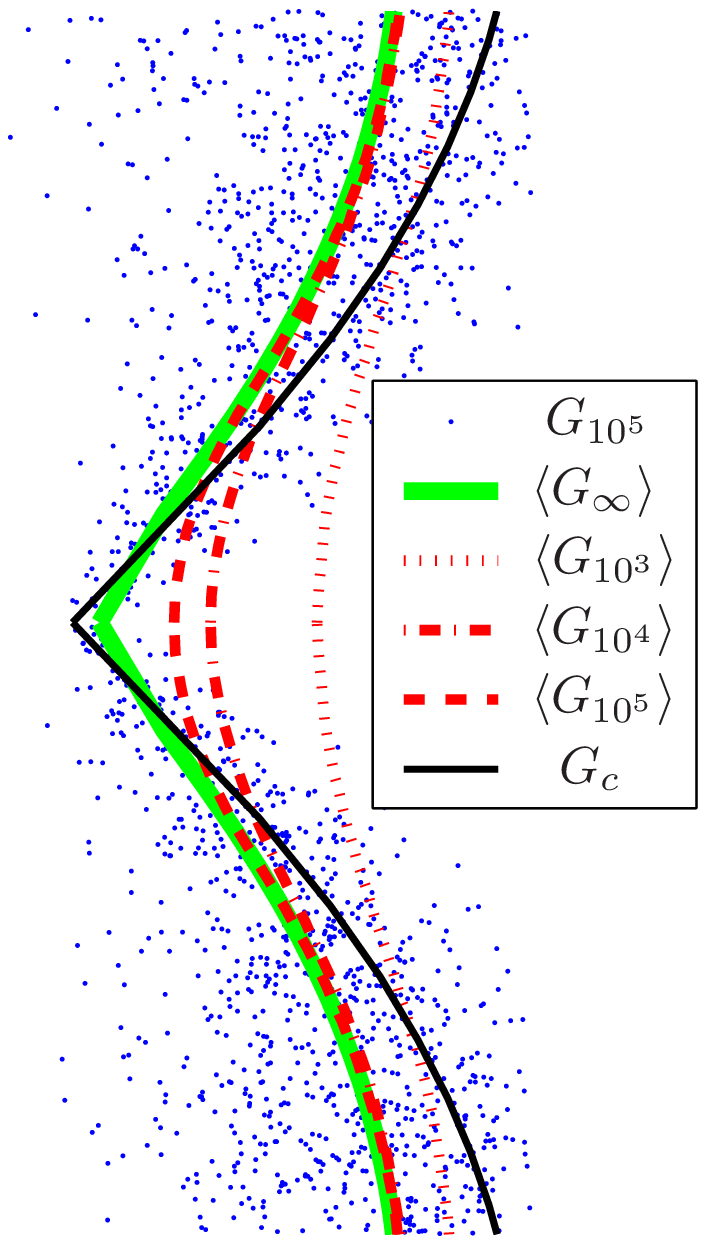}
  }
  \caption{\label{fig:cloud}
    (a) The ``cloud'' of the sampled points, $G_n(e^{i\th})$ and its
    mean field contour, $\ens{G_n(e^{i\th})}$ at $n=10^5$ (dashed)
    displayed with the asymptotic limit, $\ens{\Ginf(e^{i\th})}$
    and the steady state shape of the continuous growth, $\Gc(e^{i\th})$.
    (b) The zoom-in of the box in (a). Two contours, $\ens{G_n(e^{i\th})}$,
    at $n=10^3$ (dotted) and $10^4$ (dashed dot) are added to show the
    slow convergence to $\ens{\Ginf(e^{i\th})}$ near the rear
stagnation point.
  }
\end{figure}

Next we derive an equation for $\ens{G_n(w)}$ in the asymptotic regime.
For growing aggregates $\ld_n \to 0$ as $n \to \infty$
\cite{davidovitch99}. Following Hastings~\cite{hastings97}, we use
Eq.~(\ref{eq:HL}) to derive a linearized recursive equation for
$G_{n+1}(w)$ for $|w-e^{i\th_{n+1}}|\gg \sqrt{\ld_{n+1}}$. 
Letting $(\ld,\th)$ denote the parameters of the $(n+1)$th bump, we
obtain:
\begin{equation} \label{eq:expand}
  \begin{aligned}
    G_{n+1}&(w) \sim (1-a\ld)(G_n(w) + a\ld \Hth(w)G_n'(w)) \\
    &\sim G_n(w) + a\ld (\Hth(w)G_n'(w) - G_n(w)).
  \end{aligned}
\end{equation}
where $\Hth(w) = w\,(w+e^{i\th})\,/\,(w-e^{i\th})$
and we use $A_1(t_{n+1}) = (1+\ld)^a\; A_1(t_n)$.

Stationarity of the ensemble of rescaled clusters implies:
\begin{equation}
\ens{G_n(w)} = \intth p_\infty(\th)\ens{G_{n+1}(w)} \ .
\end{equation}
Our analysis applies for conformally invariant transport-limited
growth from an isolated seed with general angular probability
distributions, 
although we will focus on the case of ADLA, $p_\infty(\th) =
\sin(\th/2)/4$.

Using Eq.~(\ref{eq:expand}), we get the fixed-point condition:
\begin{equation} \label{eq:equib}
  \intth \,p_\infty(\th) \ens{\ld \Ginf(w)} \sim
  \intth \,p_\infty(\th) \ens{\ld \Ginf'(w)} \Hth(w).
\end{equation}
To facilitate further analysis, we approximate the left hand side
of Eq.~(\ref{eq:equib}) as
\begin{equation}
  \intth p_\infty \ens{\ld \Ginf} \approx \intth p_\infty \ens{\ld} \ens{\Ginf}
\end{equation}
and the right hand side similarly. We checked the validity of this assumption
by numerical evaluation of these two quantities for increasing values
of $n$, finding less than 1\% discrepancy for the largest clusters
($n=10^5$).
Although the stronger assumption, $\ens{\ld\Ginf}\approx
\ens{\ld}\ens{\Ginf}$, is not valid, particularly near the bump
center $w=e^{i\th}$, the correlation seems to be canceled out
in the integration along the angle.

With these assumptions, we arrive at a nonlinear integro-differential
equation for $\ens{\Ginf (w)}$, the limiting average cluster shape:
\begin{align} 
  \label{eq:Geq}
  \frac{\ens{\Ginf (w)}}{\ens{\Ginf(w)}'} &= 
  \frac{\intth\, p_\infty(\th)\,\ens{|\Ginf'(e^{i\th})|^{-2}}\, \Hth(w)}
       {\intth\, p_\infty(\th)\,  \ens{|\Ginf'(e^{i\th})|^{-2}}}  \\
  \label{eq:mf_stoc}
  &= \intth\, p_\infty(\th)\,  \LD_\infty(\th)  \, \Hth(w)
\end{align}
where we introduce a conditional probability density,
\begin{equation} \label{eq:LD_stoc}
  \LD_{n}(\th) = \frac{\ens{|G_{n}'(e^{i\th})|^{-2}}}{\intth[\th']
  p_\infty(\th')\, \ens{|G_{n}'(e^{i\th'})|^{-2}}},
\end{equation}
proportional to the average local size of a bump's pre-image
(Jacobian factor), $\ens{\ld_n}$, at angle $\th$. In
deriving Eq. ~(\ref{eq:Geq}) we assume that $\ens{A_1 (t_n)
|G_n'|^{-2}} \sim \ens{A_1(t_n)}\ens{|G_n'|^{-2}}$ as $n\to\infty$
\cite{choi05}. The stochastic nature of the aggregates is
manifested through the two different averages in
Eqs.~(\ref{eq:mf_stoc})--(\ref{eq:LD_stoc}).

\begin{figure}
  \includegraphics[width=0.7\linewidth]{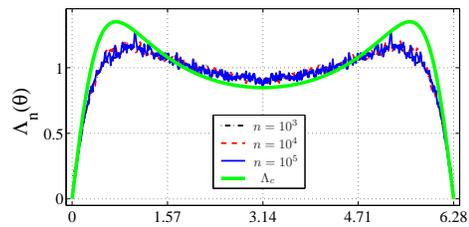}
  \caption{\label{fig:LD} 
    The angular profile of the Jacobian factor,
    $\LD_n(\th)$, which shows how the size of bump pre-images varies on
    the unit circle, at three stages of growth in simulations: $n=10^3$
    (dashed), $n=10^4$ (dash-dot) and $n=10^5$ (solid).
    The exact mean-field approximation, $\LD_c(\th)$, (thick gray)
    obtain from Eq.~(\ref{eq:sol_cont}) shows a clear difference with
    $\LD_n(\th)$.}
\end{figure}

To check the validity of Eq.~(\ref{eq:mf_stoc}), we obtain
$\LD_\infty$ from simulations and solve for $\ens{\Ginf(w)}$.
As shown in Fig.~3, the measured curves for $\LD_n(\th)$ for
$n=10^3,\,10^4$ and $10^5$ are nearly identical, so we conclude
that $\LD_{10^5}(\th)$ is a good approximation of
$\LD_\infty(\th)$.  Now we solve Eq.~(\ref{eq:mf_stoc}) by
expanding $\ens{\Ginf(w)}$ by a Laurent series and finding
recurrence relations for the coefficients, which involve integrals
of $\LD_\infty(\th)$.  We calculate 200 first coefficients and
reconstruct $\ens{\Ginf(w)}$.  The image of the unit circle
under $\ens{\Ginf(w)}$, shown in Fig.2a (thick gray), is in
excellent agreement with the converging pattern of $\ens{G_n(w)}$.

The surprising difference between the convergence rates of the
average Jacobian $\LD_n(\th)$ and the average map itself
$\ens{G_n(w)}$ is intimately related to the multifractal nature of
the distribution of the stretching factor (harmonic measure)
$|G_n'(e^{i\th})|$. Since this factor is very large around the
cusp at $\th=0$, which is dominant during the growth process,
fluctuations at the cusp do not contribute to negative moments of
the distribution of $|G_n'(e^{i\th})|$ and thus negative moments
converge much faster than positive ones, and faster than the
average map itself. Since $\LD_n(\th)$ comes from averaging
$|G_n'(e^{i\th})|^{-2}$, this observation explains its fast
convergence. This argument illustrates how the two averages
interact in Eqs.~(\ref{eq:mf_stoc})--(\ref{eq:LD_stoc}) and
suggests that the faster convergence of $\LD_n(\th)$ dominates the
morphology.

With the validity of Eq.~(\ref{eq:mf_stoc}) established, we may
consider its mean-field version, where the ensemble average is
replaced by a single conformal map, given by:
\begin{gather}
  \label{eq:mf_cont}
  \frac{\Gc(w)}{\Gc'(w)} = \intth p_\infty(\th)\, \LD_c(\th) \, \Hth(w),\\
  \label{eq:LD_cont}
  \LD_c(\th) = \frac{|\Gc'(e^{i\th})|^{-2}}{\intth[\tilde{\th}]
    p_\infty(\tilde{\th})\, |\Gc'(e^{i\tilde{\th}})|^{-2}}.
\end{gather}
Not surprisingly, the similarity solution for continuous growth,
Eq.~(\ref{eq:sol_cont}), is an exact solution of
Eqs.~(\ref{eq:mf_cont})--(\ref{eq:LD_cont}). In fact, it is possible to
derive Eqs.~(\ref{eq:mf_cont})--(\ref{eq:LD_cont}) from a different
representation of Eq.~(\ref{eq:gsb}), which has been done for the case
of DLA, $p_\infty(\th)=1/2\pi$, in Ref.~\cite{hastings97}.
Elsewhere~\cite{choi05}, we obtain an analytical form for $\LD_c(\th)$
for ADLA, which is plotted in Fig.~3 (thick gray). A small, but
significant, difference between $\LD_c(\th)$ and $\LD_\infty(\th)$ is
apparent, especially at $\th = \pi/4$ and $7\pi/4$.

The solution in Eq.~(\ref{eq:sol_cont}) can be interpreted as a
self-consistent mean-field approximation for the average conformal
map, $\ens{\Ginf(w)}$.  However, fluctuations in the ensemble
manifest themselves through the different averages in
Eqs.~(\ref{eq:mf_stoc})--(\ref{eq:LD_stoc}). As long as
$\ens{|\Ginf'(w)|^{-2}}$ is different from
$|\ens{\Ginf(w)}'|^{-2}$, $\LD_\infty(\th) \neq
\LD_c(\th)$, and thus the deviation of $\ens{\Ginf (w)}$ from
$\Gc(w)$ is inevitable.

We believe that the assumptions leading to Eq.~(\ref{eq:mf_stoc})
are quite general, and not specific to ADLA, so the continuous
dynamics should be a mean field theory (in this sense) for any
stochastic aggregation, driven by conformally invariant transport
processes, Eq.~(\ref{eq:Fi}). We conclude, therefore, that the
solution to the continuous dynamics, although similar, is not
identical to the ensemble-averaged cluster shape.  An exceptional
case is DLA in radial geometry, where isotropy implies the trivial
solution, $\ens{\Ginf(w)} = w$ and $\LD_c(\th) = 1$. Clearly,
$\Gc(w)=w$ and $\LD_c(\th)=1$ solves Eq.~(\ref{eq:mf_stoc}) with
$p_\infty(\th) = 1/2\pi$.

We expect, however, that this identity between the mean-field
approximation and the average shape of stochastic clusters will be
removed with any symmetry breaking, either in the model equations
(such as ADLA) or in the BCs (e.g.  DLA in a channel). This result
is consistent with recent simulations of DLA in a channel
geometry~\cite{somfai_ball02}, which show that the average cluster
shape, $\ens{ G_n(w)}$, is similar, but not identical, to any of
the Saffman-Taylor ``fingers'', which solve the continuous
dynamics. We expect that an analogous equation to
Eq.~(\ref{eq:mf_stoc}), relating $\ens{ \Ginf(w)} $ and
$\ens{ |\Ginf'(e^{i\th})|^{-2}}$, will hold in a channel
geometry, and Saffman-Taylor fingers should be exact solutions to
the mean-field approximation of that equation.

We conclude by emphasizing that, although Eq.~(\ref{eq:mf_stoc})
is a necessary condition for the average shape of
transport-limited aggregates in the class, Eq.~(\ref{eq:Fi}), it
does not provide a basis for complete statistical theory. Such a
theory would likely consist of an infinite set of independent
equations connecting a hierarchy of moments of the multifractal
distributions of maps $\{ \Ginf (w) \}$, and derivatives $\{
\Ginf'(w) \}$. Multifractality may speed up convergence, as
for Eq.~(\ref{eq:mf_stoc}), or slow down convergence of other
equations in this set. The mean-field approximation,
Eq.~(\ref{eq:mf_cont}), which corresponds to the continuous growth
process, can be considered as leading a hierarchy of closure
approximations to this set.

This work was supported in part by the MRSEC program of the
National Science Foundation under Award No. DMR 02-13282 (M.Z.B),
and by Harvard MRSEC (B.D).


\end{document}